\documentclass[times,doublespace]{simauth}%For paper submission

\usepackage{moreverb}
\usepackage[dvips,colorlinks,bookmarksopen,bookmarksnumbered,citecolor=red,urlcolor=red]{hyperref}
\usepackage{graphicx}
\usepackage{amsmath}
\usepackage{bbm}
\usepackage{amsfonts}
\usepackage{enumerate}
\usepackage{textcomp}
\usepackage{subfigure}
\newcommand{\bs}{\boldsymbol}

\newcommand\BibTeX{{\rmfamily B\kern-.05em \textsc{i\kern-.025em b}\kern-.08em
T\kern-.1667em\lower.7ex\hbox{E}\kern-.125emX}}

\def\volumeyear{2011}

\def\urltilda{\kern -.15em\lower .7ex\hbox{\~{}}\kern .04em}
\def\urldot{\kern -.10em.\kern -.10em}
\def\urlhttp{http\kern -.10em\lower -.1ex\hbox{:}\kern -.12em\lower 0ex\hbox{/}\kern -.18em\lower 0ex\hbox{/}}

\begin{document}

\runninghead{B.~Zhu, D.~B.~Dunson and A.~E.~Ashley-Koch}

\title{Adverse Subpopulation Regression for Multivariate Outcomes with High-Dimensional Predictors}

\author{Bin Zhu\affil{a}\affil{b}\corrauth\ , David B. Dunson\affil{a} and Allison E. Ashley-Koch\affil{b}}

\address{\affilnum{a} Department of Statistical Science, Duke University, Durham,  North Carolina 27708, U.S.A.\\
\affilnum{b} Center for Human Genetics, Duke University Medical Center, Durham, North Carolina 27710, U.S.A. 
}

\corraddr{ Bin Zhu,  Center for Human Genetics, Duke University Medical Center, Durham, North Carolina 27710, U.S.A.  E-mail:bin.zhu@duke.edu}

\begin{abstract}
Biomedical studies have a common interest in assessing relationships between multiple related health outcomes and high-dimensional predictors.  For example, in reproductive epidemiology, one may collect pregnancy outcomes such as length of gestation and birth weight and predictors such as single nucleotide polymorphisms in multiple candidate genes and environmental exposures.  In such settings, there is a need for simple yet flexible methods for selecting true predictors of adverse health responses from a high-dimensional set of candidate predictors.  To address this problem, one may either consider linear regression models for the continuous outcomes or convert these outcomes into binary indicators of adverse responses using pre-defined cutoffs.  The former strategy has the disadvantage of often leading to a poorly fitting model that does not predict risk well, while the latter approach can be very sensitive to the cutoff choice.  As a simple yet flexible alternative, we propose a method for adverse subpopulation regression (ASPR), which relies on a two component latent class model, with the dominant component corresponding to (presumed) healthy individuals and the risk of falling in the minority component characterized via a logistic regression.  The logistic regression model is designed to accommodate high-dimensional predictors, as occur in studies with a large number of gene by environment interactions, through use of a flexible nonparametric multiple shrinkage approach.  The Gibbs sampler is developed for posterior computation.   The methods are evaluated using simulation studies and applied to a genetic epidemiology study of pregnancy outcomes.
\end{abstract}


\maketitle

\section{Introduction}
\label{sec:intro}

%\emph{Background/goal}
Biomedical studies routinely collect multiple { quantitative} health outcomes and investigate how the risk of having adverse values for these outcomes is associated with predictors. The typical approach in such setting is to 1)  {   use multivariate normal linear regression in which the mean of the response distribution varies linearly with predictors; 2) first categorize the responses based on pre-specified cutoffs and then fit an ordinal logistic regression.}  The former approach is insufficiently flexible to accommodate settings in which the predictors do not simply shift the response by a fixed amount for all individuals, while the latter approach is extremely sensitive to cut-point choices.  In this article, we propose a simple alternative approach for adverse subpopulation regression (ASPR) relying on a two component mixture model that incorporates a logistic regression for the risk of falling into the minority component in the mixture, with the logistic regression model accommodating high-dimensional predictors.  {  We focus on the case when researchers are interested in dichotomizing the subjects into two classes: healthy and unhealthy group (corresponding to the majority and minority of the population); and each component is modeled by the multivariate normal distribution whose mean vector and covariance matrix change with latent class membership. This model is purposefully chosen to be simple to facilitate analyses and interpretations in settings involving high-dimensional predictors, though generalizations to multiple latent classes is straightforward as discussed in Section \ref{sec:Disc}. } 

Our approach is motivated by the Healthy Pregnancy, Healthy Baby (HPHB) Study, a prospective cohort study of pregnant women residing within Durham County, NC with the goal of identifying environmental, social and genetic factors that contribute to racial disparities in birth outcomes \cite{miranda2009environmental}. Here we focus on assessing how predictors - a large number of maternal candidate gene single nucleotide polymorphisms (SNPs), environmental exposures, and their interactions - impact the risk of low values of infant birth weight and gestational age at delivery. Such research questions cannot be addressed by the standard linear regression with continuous responses, where one models the predictor effects on the response means. The standard approach is instead to dichotomize the quantitative outcomes into binary indicators, such as low birth weight (LBW,  birth weight$<$2500g), preterm birth (PTB, gestational age at delivery$<$37weeks) and small for gestational age (SGA, birth weight less than $10$th percentile for that gestational age), and then apply logistic regression. While such analyses are easily implemented, they rely on pre-defining thresholds with the analysis results varying significantly according to the threshold choice \cite{boucher1998statistical}. 

%\emph{Proposed method and features}
We propose an alternative method for adverse subpopulation regression, which relies on a two component latent class model \cite{clogg1995latent, lindsay1995mixture, mclachlan2000finite}, with the component weights dependent on predictors via logistic regression. Related approaches are considered by Gage \cite{gage2003classification}, Gage et al. \cite{gage2008modeling} and  Schwartz et al. \cite{schwartz2010joint},  but they focused on models with fixed component weights and with the means varying with predictors. In addition, our emphasis is on applications involving high-dimensional predictors in which maximum likelihood can be expected to have poor performance.  

In such settings, it has become quite common to rely on either Bayesian methods or penalized likelihoods with penalties incorporated to favor having many coefficients estimated at or near zero, leading to variable selection and an effectively lower dimensional model.  In linear regression and generalized linear models, such methods have become standard, with the Lasso \cite{tibshirani1996regression}, elastic net \cite{zou2005regularization} and relevance vector machine \cite{Tipping2001} providing popular examples.  The penalized likelihood estimators have a Bayesian interpretation in corresponding to the mode of the posterior distribution obtained under carefully chosen priors on the coefficients, with the Laplace leading to the Lasso \cite{park2008bayesian} and a $t$-distribution with low degrees of freedom leading to the relevance vector machine. MacLehose and Dunson \cite{maclehose2010bayesian} recently proposed a new class of multiple shrinkage priors that allow shrinkage towards not only zero but also other values, leading to improved performance in estimating non-zero coefficients.  We will consider these and other shrinkage priors in the context of our ASPR model.

%\emph{Organizations}
The remainder of the paper is organized as follows. In Section \ref{sec:SBLCM}, we introduce the proposed Bayesian adverse subpopulation regression model, describing both fully Bayes and fast two-stage approaches for inference. Section \ref{sec:PostCpt} provides details of an Markov chain Monte Carlo (MCMC) algorithm. Section \ref{sec:simu} presents the simulation results, evaluating and comparing the proposed methods with existing methods. In Section \ref{sec:app}, we apply the model to pregnancy outcome data. The article concludes with a discussion in Section \ref{sec:Disc}.  

\section{Bayesian Adverse Subpopulation Regression}
\label{sec:SBLCM}
\subsection{Model Formulation}
\label{sec:formulation}
Suppose we collect the data $(\bs{y}^\prime_i, \bs{x}^\prime_i)$ for subject $i$, $i=1,2,\dots,n$, where $\bs{y}_i$ is an $s \times 1$ vector of outcomes and $\bs{x}_i$ is a $p \times 1$ vector of predictors.  We make the simplifying assumption that there are two types of individuals, with $z_i=0$ denoting healthy individuals and $z_i=1$ for potentially unhealthy individuals. In addition, we assume for identifiability that the unhealthy individuals are in the minority, with the specific constraints and prior information included for identifiability discussed in detail in Section \ref{subsec:prior}. This is a simplification which is made for ease in interpretation, assessment of risk, and scaling to higher dimensions while accommodating the curse of dimensionality that arises. In many cases, such a simplification is made in advance of the analysis by taking one or more response variables and defining cutoffs to dichotomize the data prior to analysis.  However, it is well known that results are quite sensitive to the choice of cutoff \cite{greenland1995avoiding}, and hence we prefer allowing $z_i$ to be an adverse health status latent variable.  By using a Bayesian approach, we can fully accommodate uncertainty in imputing $z_i$ and avoid forcing any hard threshold on the observed quantitative traits.

Denote $\omega_1(\bs{x}_i)=\textsf{Pr}(z_i=1 \mid \bs{x}_i)$ as the probability of allocating subject $i$ to the unhealthy population and let $\omega_2(\bs{x}_i)=1-\omega_1(\bs{x}_i)$. We then express the conditional density of the response $\bs{y}_i$ given predictors $\bs{x}_i$ as 
\begin{equation}
\label{eq:aspr_1}
f(\bs{y}_i \mid \bs{x}_i) = \sum_{h=1}^{2}  \omega_h(\bs{x}_i)\textsf{N}_s(\bs{y}_i \mid \bs{\theta}_h, \bs{\Sigma}_h),
\end{equation}
where $\textsf{N}_s(\bs{y}_i \mid \bs{\theta}_h, \bs{\Sigma}_h)$ is the $s$-dimensional Normal distribution with mean vector $\bs{\theta}_h$ and covariance matrix $\bs{\Sigma}_h$. 
We assume that variability in the healthy and unhealthy groups in the quantitative traits is adequately characterized by a multivariate normal distribution for the sake of identifiability.  Potentially, one could instead use a heavier-tailed distribution such as a multivariate $t$-distribution, though allowing outlying individuals in the two groups may degrade separation of the two components.  For example, an unhealthy individual may be in the tails of the $t$-distribution for the healthy component.
In addition, $\omega_1(\bs{x}_i)$ in expression \eqref{eq:aspr_1} depends on predictors $\bs{x}_i$ through a logistic regression model: 
\begin{equation}
\label{eq:aspr_2}
\omega_1(\bs{x}_i) = \frac{\exp(\gamma+\bs{x}^\prime_i\bs{\beta})}{1+ \exp(\gamma+\bs{x}^\prime_i\bs{\beta})},
\end{equation}
where the coefficient vector $\bs{\beta}=(\beta_{1},\beta_{2},\dots,\beta_{p})^\prime$ characterizes the effect of predictors on the risk of falling in the minority subpopulation.  Due to the logistic regression form, the exponentiated $\beta_j$ coefficients can be  interpreted as odds ratios.

\subsection{Prior Specification}
\label{subsec:prior}
For the ASPR model in \eqref{eq:aspr_1}-\eqref{eq:aspr_2}, identifiability of the unhealthy subpopulation necessarily relies on prior information. In the absence of some prior knowledge,  the two subgroups would be exchangeable,  and we would encounter a label-ambiguity problem.  Removal of this problem through appropriate priors is one of the advantages of the simple two component framework over more complex latent class regression models having unknown numbers of components. The most common approach would place restrictions on the means of the components; for example, ordering the components in advance by letting $\theta_{11}<\theta_{21}$. This approach assumes that low values of the first response variable are adverse, which may be reasonable for a given study but is not so in general. Moreover, placing restriction on the means will fail to solve the label ambiguity problem if the components are not separated sufficiently.

Thus, we consider alternative strategies depending on the application. The first is to elicit informative values for the means and covariance matrix in the two components from prior empirical knowledge of the typical distribution of the responses in healthy and unhealthy groups. In the absence of such extra knowledge, we may fit a mixture of two multivariate normals to the data using EM for maximum likelihood estimation, defining the minority component to be adverse.  This can be done either from historical data or the current data.  Then, fix the $\bs{\theta}_h$, $\bs{\Sigma}_h$ at these estimates in the subsequent analysis.  This runs the risk of under-estimating uncertainty but has the advantage of simplifying interpretation and completely eliminating identifiability concerns. 

Another alternative is to  specify conditionally conjugate prior distributions for $\bs{\theta}_h$, $\bs{\Sigma}_h$ and $\gamma$ as follows, 
\begin{equation*}
(\bs{\theta}_h, \bs{\Sigma}_h) \sim \textsf{NIW}_s(\bs{\theta}_h, \bs{\Sigma}_h \mid \bs{\theta}_0, \psi_0, \rho_0, \bs{\Sigma}_0),  \quad h=1,2,
\end{equation*}
where $\textsf{NIW}_s(\bs{\theta}_h, \bs{\Sigma}_h \mid \bs{\theta}_0, \psi_0, \rho_0, \bs{\Sigma}_0)$ is the Normal-inverse-Wishart distribution proportional to $|\bs{\Sigma}_h|^{-(s+\rho_0+2)/2}\exp\left\{-\frac{1}{2}\text{tr}(\bs{\Sigma}_0\bs{\Sigma}_h^{-1})-\frac{\psi_0}{2}(\bs{\theta}_h-\bs{\theta}_0)^\prime\bs{\Sigma}^{-1}(\bs{\theta}_h-\bs{\theta}_0) \right\}$. As weakly informative empirical Bayes priors, the hyperparameters $\bs{\theta}_0$ and $\bs{\Sigma}_0$ are chosen to be the sample means and  covariance matrix for all subjects, and we set $\psi_0=1$ and $\rho_0=s+2$ to reduce the prior information. Additionally, we place an informative prior on the intercept in the logistic regression model \eqref{eq:aspr_2} after centering the predictors, $\gamma \sim \textsf{N}_1(\gamma \mid \gamma_0, \lambda_0)$. For example, by choosing $\gamma_0=-2.20$ and $\lambda_0=2.42$ for the intercept,  the expected baseline probability (\emph{a priori}) of an adverse response is $10\%$ and falls in the range between $3\%$ and $30\%$ with $0.95$ probability.

As for the priors for $\bs{\beta}$, if $\bs{x}_i$ is low-dimensional,  we can rely on standard choices, such as independent Gaussian distributions with modest variance.  However, as the number of predictors increases, we need some approach for addressing the high dimensionality.  A common strategy in the frequentist literature is to use sparse penalized regression (e.g., Lasso, elastic net, etc) to favor many elements of $\bs{\beta}$ that are equal to zero while shrinking the non-zero elements toward zero. In the Bayesian literature, a rich variety of shrinkage priors have been proposed for high-dimensional regression coefficients, with most approaches relying on priors that are centered at zero, potentially with a point mass incorporated to allow variable selection. Hierarchical shrinkage priors that are centered at zero can potentially lead to over-shrinkage of coefficients that are not close to zero. Such over-shrinkage can be reduced by choosing a prior which is concentrated near zero with very heavy tails, but in that case there is no borrowing of information or incorporation of prior knowledge in estimating the coefficients that are not close to zero. 

As an alternative approach that had excellent performance in high-dimensional logistic regression, MacLehose and Dunson \cite{maclehose2010bayesian} proposed a multiple shrinkage prior (MSP) $\beta_j \sim \int \textsf{DE}(\beta_j \mid \mu_j, \tau_j) {d} \textsf{P}(\mu_j, \tau_j)$, where $\textsf{DE}(\beta_j \mid \mu_j, \tau_j)$ is the double exponential (Laplace) distribution with location parameter $\mu_j$ and scale parameter $\tau_j$;  the mixture distribution $\textsf{P}$ is assigned a modified Dirichlet process prior that incorporates a mass at $\mu_j=0$ for the first component. In stick-breaking form \cite{sethuraman1994constructive}, this prior for $\textsf{P}(\cdot)$ can be expressed as
\begin{align*}
\textsf{P}(\cdot) &=\pi_1 \delta_{(0,\tau^{*}_{1})}(\cdot) + \sum_{t=2}^{\infty}\pi_t\delta_{(\mu_{t}^{*},\tau^{*}_{t})}(\cdot),\\
%\label{eq:MSP4}
\mu_{t}^{*} &\sim \textsf{N}_1(\mu_{t}^{*} \mid c, d ),\\
 \tau^{*}_{1} &\sim \textsf{Gamma}(\tau^{*}_{1} \mid a_0, b_0), \quad \tau^{*}_{t} \sim \textsf{Gamma}(\tau^{*}_{t} \mid a_1, b_1)
,\\
%\label{eq:MSP5}
\pi_t &=V_t \prod_{l<t}(1-V_l), \quad V_t \sim \textsf{Beta}(1, \alpha),
\end{align*}
where $\textsf{Gamma}(\tau \mid a, b)=1/[b^a\Gamma(a)]\tau^{a-1}\exp(-\tau/b)$ with mean $a \times b$; $\delta_\theta(\cdot)$ is the probability measure with all its mass at $\theta$; following MacLehose and Dunson \cite{ maclehose2010bayesian},  we choose $c=0$, $a_0=b_0=30$, $a_1=b_1=6.5$ and $\alpha=1$. Consequently, through MSP, the coefficients $\bs{\beta}$ will be shrunk to multiple locations $\mu_{j}^{*}$s, including zero in the first cluster ($t=1$), corresponding to the usual Bayesian Lasso prior, while the other components are centered at unknown locations away from zero. {  For $\beta_j$ and $\beta_{j^\prime}$ belonging to the same cluster $t$, $\beta_j \neq \beta_{j^\prime}$ with $E(\beta_j)=E(\beta_{j^\prime})=\mu_{t}^{*}$ and $Var(\beta_j)=Var(\beta_{j^\prime})=\tau^{*2}_{t}$.  }  In our application of the ASPR model, it is unlikely that any of the predictors being considered have a log-odds ratio of falling in the adverse sub-group outside of $\beta_j \in [-1,1]$, corresponding to an interval of $[0.37, 2.72]$ for the odds ratio. In most genetic epidemiology studies involving complex health conditions, one expects at most a modest deviation from a log-odds ratio at 1.00 for single SNPs or SNP $\times$ environment interactions.  This small signal-to-noise ratio is one aspect that makes detection of important variants so challenging.  To express this prior information, while inflating the prior variance somewhat to corresponding to a ``weakly informative'' prior \cite{gelman2008weakly}, we let $d=0.1507$, which leads to $\textsf{Pr}( \mu_t^* \in
[-1,1]) = 0.99$ {\em a priori}.

% In our application of the ASPR model, it is reasonable to assume that the given predictor would have small (if there is any) influence on the probability of allocating the subject to the unhealthy group, and that $\beta_j$ would be in the range of -1 and 1(corresponding to odds ratio $\exp(-1)=0.37$ and $\exp(1)=2.72$). Hence, we choose $d=0.1507$ such that the with $99\%$ probability $\beta_j$ would be shrunk to the location $\mu_j^{*}$, which lies in $[-1,1]$. 
   
\section{Posterior Computation}
\label{sec:PostCpt}
In describing an approach for posterior computation, we focus on the approach described in Section 2 that places a normal-inverse-Wishart prior on the component-specific parameters, an informative prior on the intercept $\gamma$ for identifiability,  and a mixture of double exponential shrinkage prior on the high-dimensional vector of $\bs{\beta}$ coefficients.  This approach is straightforward to modify to accommodate the other approaches described in Section 2.  For example, to instead use the two-stage plug-in approach, we would run the EM algorithm first to estimate $\mu_h,\Sigma_h$ for $h=1,2$ and then would hold these component-specific parameters fixed in the proposed data augmentation Gibbs sampling algorithm to be described below.  In addition, if an alternative shrinkage prior were used for the coefficients $\beta_j$, then one could simply modify the sampling steps for updating the $\beta_j$ appropriately.  For scale mixture of normal priors, such as double exponentials, $t$ priors or other standard choices, this is straightforward.

If we observe the latent subpopulation index $z_i$ directly for each individual and are interested in the coefficients $\bs{\beta}$, then we could apply the MCMC algorithm of MacLehose and Dunson \cite{maclehose2010bayesian} directly.  However, because we do not observe $z_i$ for any of the subjects, we instead modify their algorithm to include steps for imputing $z_i$ from the corresponding full Bernoulli conditional posterior distribution and sampling the mean and covariance specific to each component.    
We start by relating the latent subpopulation index $z_i=I(g_i>0)$ to an auxiliary random variable $g_i$, where $I(\cdot)$ is the indicator function, which equals $1$ when $g_i>0$ and $0$ otherwise. To induce expression \eqref{eq:aspr_2} through marginalizing $g_i$, we assume $g_i$ follows a logistic distribution centered on $\gamma+\bs{x}^\prime_i\bs{\beta}$.
 {  Holmes and Held \cite{holmes2006bayesian} proposed a data augmentation MCMC algorithm for posterior computation in logistic regression models relying on characterizing the latent $g_i$ as a scale mixture of normals, with the square root of the scale parameters following a Kolmogorov-Smirnov (KS) distribution.  Due to lack of conjugacy of the conditional posteriors of scale parameters specific to each subject, they recommend using rejection sampling.  However, use of a large number of rejection sampling steps can lead to inefficiencies, so we instead apply an alternative data augmentation scheme. }  Following O'Brien and Dunson \cite{o2004bayesian}, the logistic distribution can be almost exactly approximated by a noncentral $t$-distribution $t_v( g_i \mid \gamma+\bs{x}^\prime_i\bs{\beta}, \sigma^2)=\int_0^{\infty} \textsf{N}_1(g_i \mid \gamma+\bs{x}^\prime_i\bs{\beta}, \sigma^2/\phi_i)\textsf{Gamma}(\phi_i \mid \nu/2, 2/\nu)d\phi_i$, when we set $\sigma^2=\pi^2 (\nu-2)/2\nu$ with degree of freedom $\nu=7.3$.  {  Kinney and Dunson \cite{kinney2007fixed} showed that posterior distributions of $g_i$ estimated with the Holmes and Held  \cite{holmes2006bayesian} and O'Brien and Dunson \cite{o2004bayesian} algorithms are  essentially completely indistinguishable given sufficient numbers of MCMC samples.}  We outline the Gibbs sampler for Bayesian adverse subpopulation regression in the following steps:
\begin{enumerate}[(a)]
\item Draw $\bs{\theta}_h$ and $\bs{\Sigma}_h$ from  
$\textsf{NIW}_s(\bs{\theta}_h, \bs{\Sigma}_h \mid \bs{\hat{\theta}}_h, \hat{\psi}_h, \hat{\rho}_h, \bs{\hat{\Sigma}}_h), h=1,2,$ where
\begin{align*}
\hat{\bs{\theta}}_h &= \frac{n_h}{n_h+\psi_0}\bs{\bar{y}}_h+\frac{\psi_0}{n_h+\psi_0}\bs{\theta}_0,\\
\hat{\psi}_h &= n_h + \psi_0,\\
\hat{\rho}_h &= n_h + \rho_0,\\
\bs{\hat{\Sigma}}_h &= \bs{{\Sigma}}_0 + \bs{S}_h + \frac{n_h}{1+n_h\psi^{-1}_0}(\bs{\bar{y}}_h-\bs{\theta}_0)(\bs{\bar{y}}_h-\bs{\theta}_0)^\prime,
\end{align*} 
with $n_1 = \sum_{i=1}^n I(z_i=1)$, $\bs{\bar{y}}_1 = \frac{1}{n_1} \sum_{i:z_i=1} \bs{{y}}_i$ and 
$\bs{S}_1 = \sum_{i:z_i=1}(\bs{{y}}_i-\bs{\bar{y}}_1)(\bs{{y}}_i-\bs{\bar{y}}_1)^\prime$; $n_2 = \sum_{i=1}^n I(z_i=0)$, $\bs{\bar{y}}_2 = \frac{1}{n_2} \sum_{i:z_i=0} \bs{{y}}_i$ and 
$\bs{S}_2 = \sum_{i:z_i=0}(\bs{{y}}_i-\bs{\bar{y}}_2)(\bs{{y}}_i-\bs{\bar{y}}_2)^\prime$.

\item Impute component indicator $z_i$ from the conditional Bernoulli distribution by setting $z_i=1$ with probability $\frac{\omega_1(\bs{x}_i)\textsf{N}_s(\bs{y}_i \mid \bs{\theta}_1, \bs{\Sigma}_1)}{\sum_{h=1}^{2}  \omega_h(\bs{x}_i)\textsf{N}_s(\bs{y}_i \mid \bs{\theta}_h, \bs{\Sigma}_h)}$ for $i=1,2,\dots,n$.

\item Augment auxiliary variable $g_i$, $i=1,2,\dots,n$, sampled from the  normal distribution $\textsf{N}_1(g_i\mid \gamma+\bs{x}^\prime_i\bs{\beta}, \sigma^2/\phi_i )$, which is truncated above (below) by zero when $z_i =0$ ($z_i=1$).

\item Update $\phi_i$ from $\textsf{Gamma}\left(\phi_i \mid \frac{\nu+1}{2}, \frac{2}{\nu+(g_i-\gamma-\bs{x}^\prime_i\bs{\beta})^2/\sigma^2} \right)$, $i=1,2,\dots,n$.

\item Update the regression coefficients $\bs{\beta}^*=(\gamma, \bs{\beta}^\prime)^\prime$ given $g_i$, $\phi_i$ and other parameters, following the MCMC algorithm of MacLehose and Dunson \cite{maclehose2010bayesian}. 
\end{enumerate}
Although we illustrate the algorithm focusing on the multiple shrinkage prior, the above algorithm could be easily modified for different shrinkage priors of $\bs{\beta}$ by using the corresponding sampling algorithm in the step (e). Moreover, we could combine the shrinkage and selection method (e.g. Lasso and elastic net) for logistic regression with step (a) and (b) to get a Monte Carlo EM algorithm \cite{wei1990monte} for the adverse subpopulation regression.

\section{Simulations}
\label{sec:simu}
In this section, we examine the performance of our approach along with alternative simple two-stage methods through simulation studies. In the two-stage methods, the binary indicators for the adverse subpopulation are generated in the first stage. They are chosen as the true binary indicators (known for simulation data), the maximum a posteriori allocation from a simple two component mixture model estimated via maximum likelihood implemented with the EM algorithm, or values obtained using prespecified cutoffs. In the second stage, we fit both the standard logistic regression model without penalization and  the penalized logistic regression models with the shrinkage methods Lasso and elastic net \cite{friedman2010reg}.

One hundred datasets were simulated to represent the data observed in the HPHB data set.  In particular, we simulated 813 women with two response variables corresponding to infant birth weight and gestational age at delivery.  We used maternal genotype for 100 SNPs as predictors that were fixed across the simulations, with only the response variable generation varying.  By using the real SNP data, we obtained simulated datasets with a realistic dependence structure among the predictors, which is important given that the dependence structure can have a fundamental impact on variable selection and estimation performance.  We simulated data under the model proposed in Section \ref{sec:formulation}, with $\bs{\beta}$ chosen so that the first ten elements were set equal to 0.8 (corresponding to odds ratios for the minor allele of $\exp(0.8)=2.23$) and the remaining elements were set equal to zero (corresponding to no association with risk of falling in the adverse subpopulation for SNPs $11,\ldots, 100$).  We simulated  $\bs{y}_i$ based on expression \eqref{eq:aspr_1}, where the $\bs{\theta}_h$ and $\bs{\Sigma}_h$ were set equal to the maximum likelihood estimates from the HPHB dataset by using a  two component latent class model without predictors.   
    
We applied the proposed ASPR model with default priors specified in Section \ref{subsec:prior} and the two-stage methods to the simulated datasets. 
For ASPR model, we implemented the data augmentation Gibbs sampling algorithm outlined in Section \ref{sec:PostCpt}. The sampling ran for 11,000 iterations, 1,000 iterations were discarded as a burn-in and every 10th sample was saved to thin the chain. The trace and autocorrelation plots of the posterior samples were examined to determine the convergence. We used the cyclical coordinate descent algorithm by Friedman et al. \cite{friedman2010reg} to find the Lasso or elastic net regularization paths
for penalized logistic regression models. Ten-fold cross-validations were used to select the optimal shrinkage parameter which gave the minimal deviance.

\begin{table}
\caption{Simulation results to compare coefficient estimation and variable selection by the ASPR model, the standard logistic regression models without penalization and the penalized standard logistic regression models with shrinkage methods. \label{tbl:simu}}
\centering
{\footnotesize
\begin{tabular}{ccccc}
\toprule
&ASPR-MSP&Truth&Classification&Cutoff\\
\midrule
\\
&&\multicolumn{3}{c}{Logit-standard}\\
\cline{3-5}
MSE&&&&\\
(Nonnull/Null)&0.220/0.008&0.582/0.674&1.170/0.886&0.274/0.471\\
interval.length&&&&\\
(Nonnull/Null)&0.983/0.439&1.802/3.401&1.842/5.651&1.422/3.283\\
TPR&0.286&0.683&0.570&0.525\\
FPR&0.010&0.163&0.168&0.144\\
AUC &0.926&0.809&0.742&0.731\\
\\
&&\multicolumn{3}{c}{Logit-Lasso}\\
\cline{3-5}
MSE&&&&\\
(Nonnull/Null)&&0.274/0.005&0.513/0.004&0.381/0.003\\
TPR&&0.790&0.701&0.646\\
FPR&&0.141&0.114&0.097\\
AUC&&0.866&0.821&0.794\\
\\
&&\multicolumn{3}{c}{Logit-ElasticNet}\\
\cline{3-5}
MSE&&&&\\
(Nonnull/Null)&&0.233/0.005&0.470/0.004&0.359/0.003\\
TPR&&0.905&0.838&0.771\\
FPR&&0.242&0.210&0.175\\
AUC &&0.919&0.876&0.845\\
%&\\
%&&&&&&&&&&\\
%AUC&&&&&&&&&&\\
%Mean&0.270/0.003&&0.279/0.005& 0.563/0.005&0.513/0.001&& 0.239/0.006&0.503/0.006&0.516/0.001\\
%Min&0.105/0.000&&0.102/0.000& 0.512/0.000&0.351/0.000&& 0.064/0.000&0.445/0.000&0.375/0.000\\
%Max&0.483/0.017&&0.474/0.017&0.624/0.017&0.615/0.005&& 0.472/0.016&0.583/0.015&0.619/0.004\\
\bottomrule
\end{tabular}
}
\end{table} 

%&\multicolumn{1}{c}{ASPR-MSP}  & \multicolumn{3}{c}{Logit-Naive}  &\multicolumn{3}{c}{Logit-Lasso}  & \multicolumn{3}{c}{Logit-ElasticNet}
%\\

\begin{figure}[h!]
  \centering
    \includegraphics[width=0.9\textwidth,angle=270]{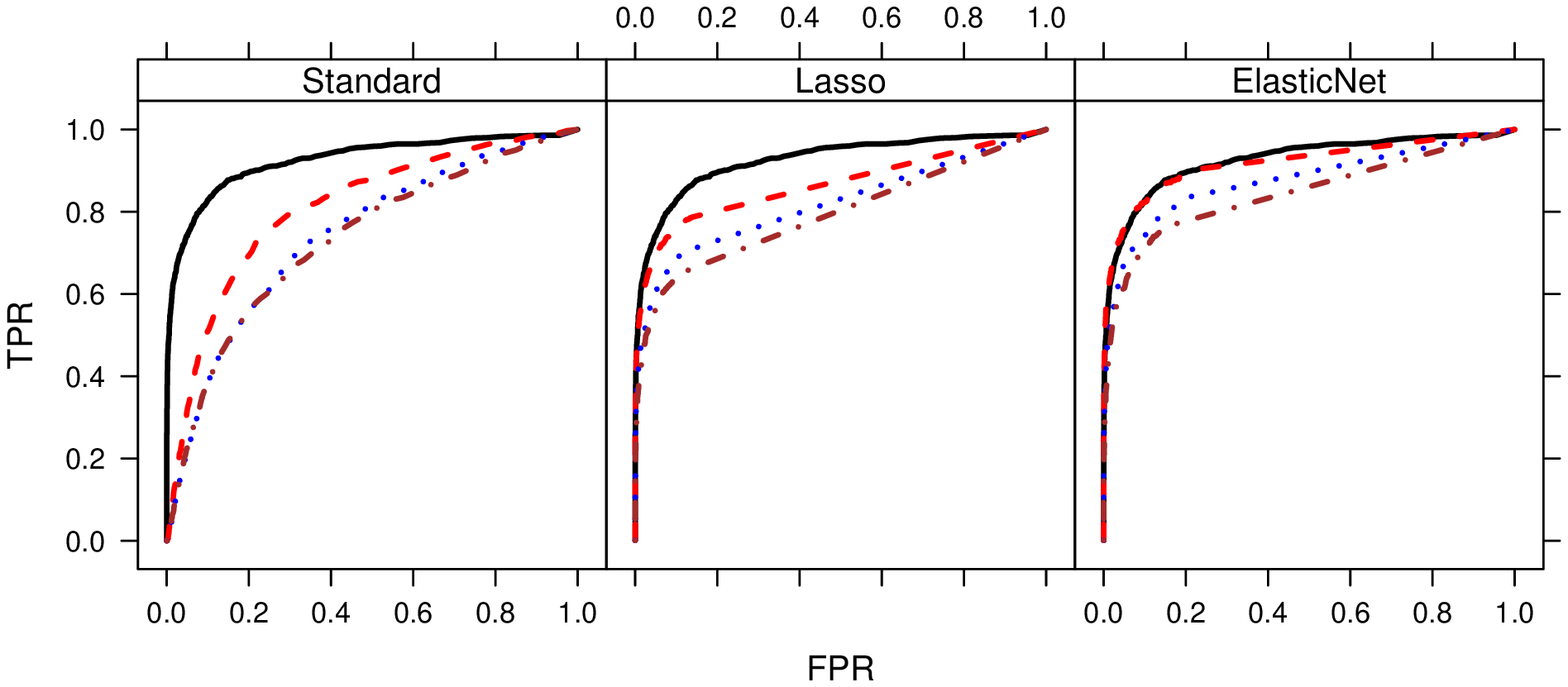}
  \caption{Receiver Operating Characteristic (ROC) curves for different methods: \textemdash\textemdash \; by ASPR-MSP, $---$ by the two-stage methods with true indicators,  $\cdotp \cdotp \cdotp$  by the two-stage methods with subjects allocated by maximum a posteriori,  $-\; \cdotp -$ by the two-stage methods with indicators defined by cutoffs. \label{fig:roc}}
\end{figure}

We first compared the estimation performance measured by mean squared errors (MSEs) which were calculated for each coefficient across 100 datasets. Table \ref{tbl:simu} presents the averaged MSEs obtained across the first ten non-null coefficients and the remaining ninety null coefficients. The averaged MSEs of non-null coefficients by ASPR model are smaller than those given by the two-stage methods even with true indicators. More importantly, when the true indicators are unknown, a common scenario in practice, the two-stage methods rely on indicators generated either by classification algorithm(here the maximum a posteriori allocation) or by medical cutoffs will inflate the MSEs of non-null coefficients significantly. This observation indicates that if part of subjects are mis-classified in the two-stage methods, the coefficient estimates would be affected in the standard and penalized logistic regressions. A better solution is thus to avoid the classification or using cutoffs in the first place. In addition to demonstrating smaller MSEs, for ASPR model we can easily obtain the credible intervals of coefficient based on posterior samples. As shown by Table \ref{tbl:simu}, the averaged lengths of $90\%$ credible interval by ASPR model is much narrower than $90\%$ confidence intervals by standard logistic regression models for both the non-null and null coefficients.           
          
We also compared the variable selection ability for different methods. The predictor $\beta_j$ is selected if its $90\%$ credible or confidence interval dos not contain zero for the ASPR model and standard logistic regression model, or if estimate of $\beta_j$ is not equal to the zero for the penalized logistic regression models. The true positive rate (TPR) and false positive rate (FPR) in selection are presented in Table \ref{tbl:simu}. Compared to the other methods, the ASPR model achieves both lower TPR and FPR, which can be increased if we use the narrower credible interval. Although the low TPR in ASPR model is undesirable, the low FPR may be beneficial for the study in which the false selection of predictors will lead to serious consequence. {  In applications involving thousands to hundreds of thousands of genetic markers, it is critical to control the FDR at a low level to avoid producing hundreds of false findings.} Based on the substantive knowledge, the predictor $\beta_j$ may also be selected if $|\beta_j| > \epsilon$. Figure \ref{fig:roc} illustrates the receiver operating characteristic (ROC) curve, which plots the TPRs and FPRs as $\epsilon$ is varied. It it clear that the ROC curve by ASPR model is closer to the top left corner with higher value of area under the curve (AUC, shown in Table \ref{tbl:simu}), which suggests a better trade-off between the TPR and FPR in terms of variable selection.     

\section{Application to Pregnancy Outcomes}
\label{sec:app}
There is increasing appreciation that interactions between the genetic and environmental factors contribute to adverse birth outcomes. In this analysis,  we investigated the effects of maternal genotype and their interaction with lead and tobacco exposure on adverse birth outcomes in the infant, adjusting for several confounding factors. The dataset included 813 non-Hispanic black pregnant women who had singleton pregnancy and were less than 28 weeks gestation at the time of enrollment in HPHB study. Based on published studies, we focused on 31 candidate genes which are involved with maladaptive inflammatory regulation, maternal-fetal circulation, stress response, and environmental contaminant metabolism. For those candidate genes, we selected 275 haplotype tagging SNPs which effectively capture the genetic diversity of these genes. Please see Swamy et al. \cite{swamy2010} and Ashley-Koch et al. \cite{Ashley2011} for further details on genotyping approaches. A detailed description of the SNPs and genes used in this analysis can be found in the Web Appendix.  For the purpose of this analysis, we assumed that the risk for adverse birth outcomes would be associated with minor alleles. The value of each SNP was recorded as one if the mother carried the less frequent allele and as zero otherwise. In addition to the genetic data, we measured maternal blood levels of lead and cadmium. The interaction of lead and cadmium with the SNPs in relation to gestational age and birth weight is an important research question.
We also controlled confounding factors by including them in the analysis. These confounders are mother's age, recorded as age group 18-20, 21-35 vs 35+; education, as no college vs some college; insurance, as private vs others; parity,  as zero vs others; infant sex, as male vs female.      

\begin{figure}
  \centering
\subfigure[]{\includegraphics[width=0.48\textwidth]{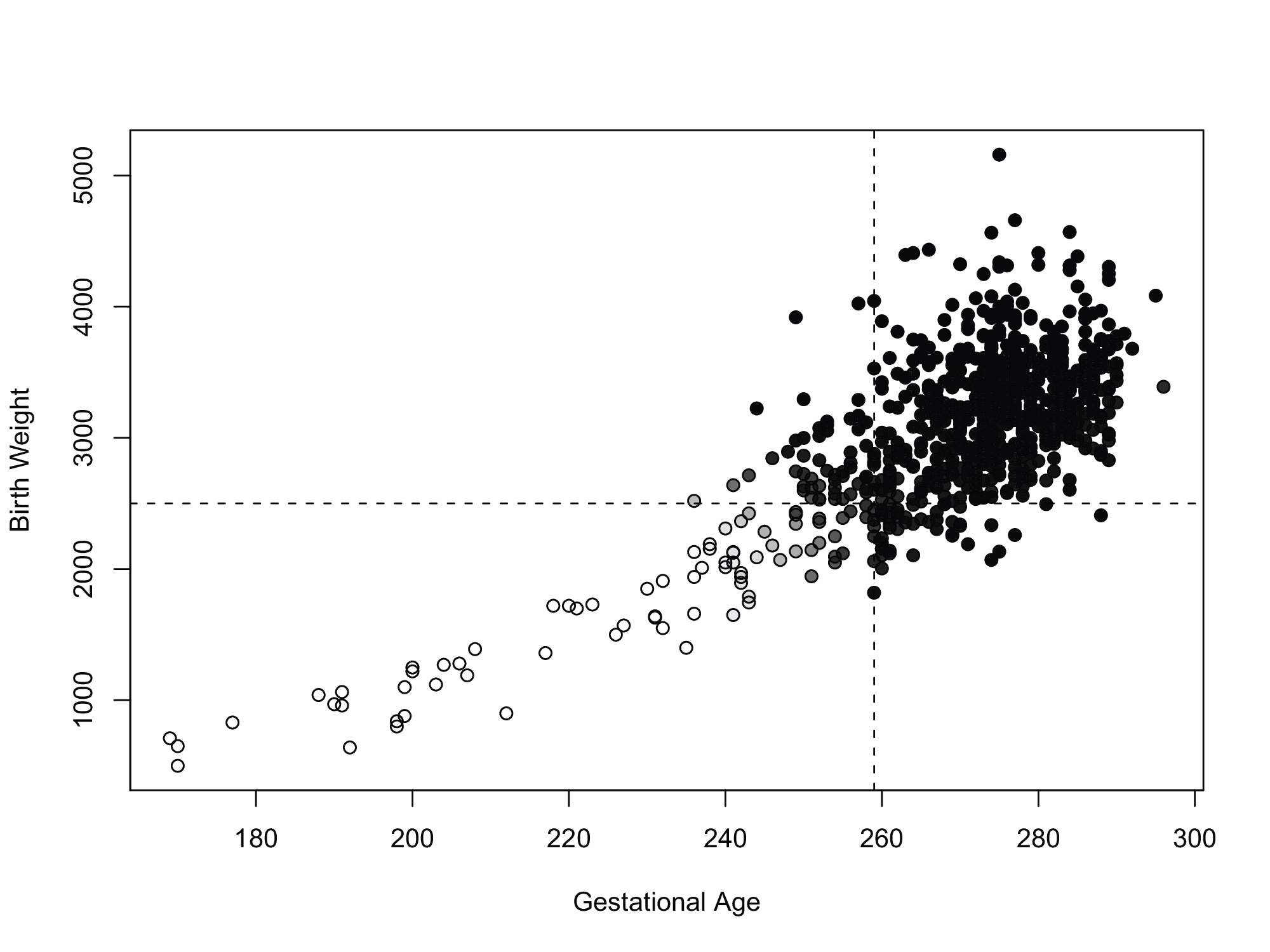}\label{fig:scplot}}
\subfigure[]{\includegraphics[width=0.48\textwidth]{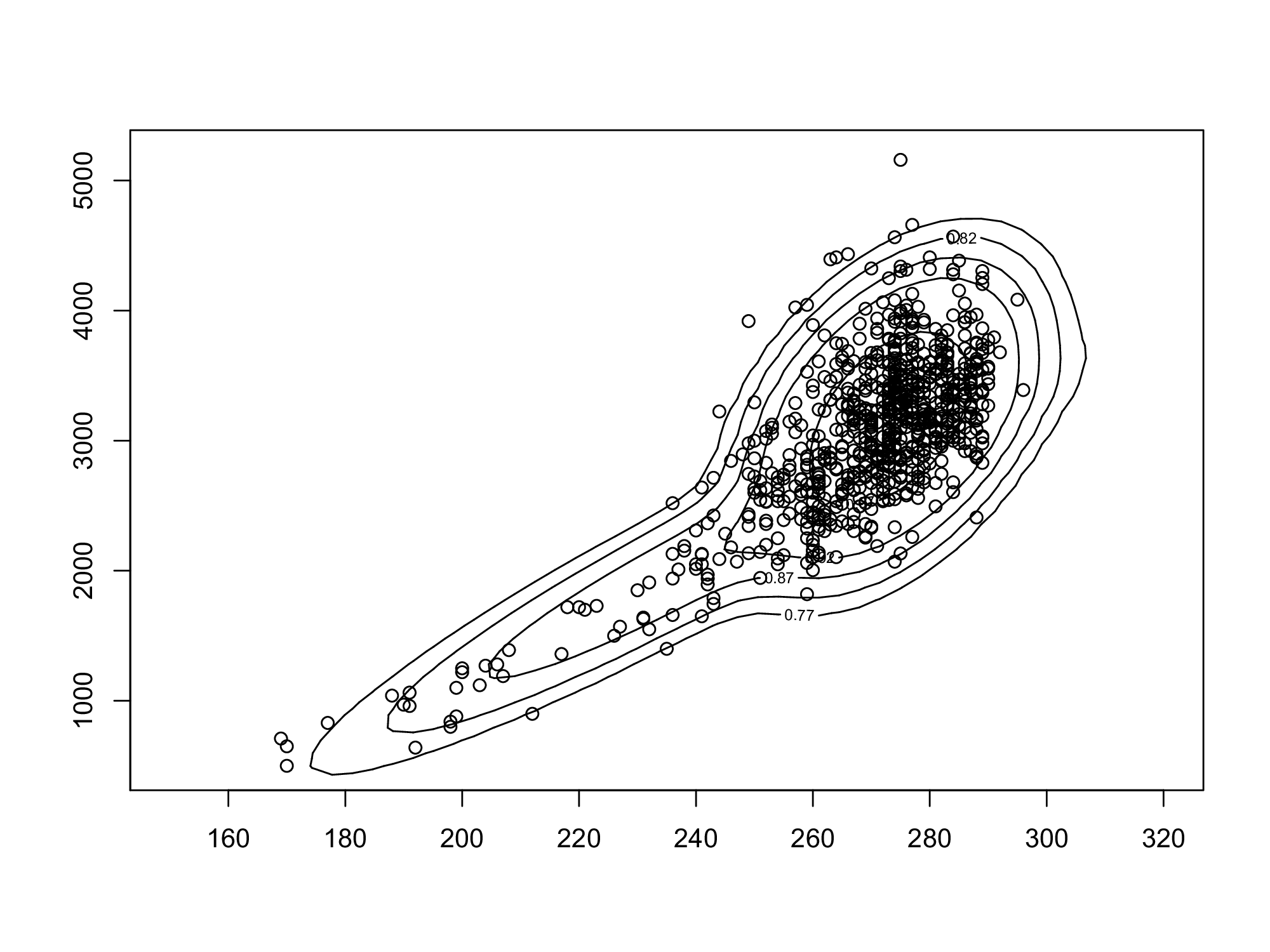}\label{fig:post.pred}}
  \caption{ Scatter plots of birth weight (grams) and gestational age (days) overlaid with (a) posterior mean of allocation weight $\omega_2(\bs{x}_i)$ with vertical dash line at 259 and horizontal dash line at 2500; (b) posterior predictive density of observations ($\circ$).}
\end{figure}

We fit the ASPR model with default priors and ran the MCMC algorithm for 11,000 iterations with the first 1,000 iterations discarded as burn-in and every 10th remaining draw retained for analysis. The trace plots and the autocorrelation plots suggested the algorithm converged fast and mixes well. Table \ref{tbl:post.comp} presents the posterior summary for the component parameter $\bs{\theta}_h=(\theta_{h1}, \theta_{h2})^\prime$ and
$\bs{\Sigma}_h=  \left( \begin{array}{cc}
\Sigma_{h11} & \Sigma_{h12}  \\
\Sigma_{h21} & \Sigma_{h22}  \\
\end{array} \right) 
$. The table indicates that the healthy group in general has longer gestational age with higher birth weight, compared to the unhealthy group. In addition, the subjects in the healthy group are more homogeneous with the smaller values in the components of $\bs{\Sigma}_h$. Figure \ref{fig:scplot} shows shaded circles at the raw data points, with the darkness of the shading being proportional to the estimated posterior probability of allocation to the healthy subgroup.  Standard cutoffs for defining preterm birth and low birth weight are also shown.  Although most of the children that are in the preterm and low birth weight bin have small posterior probabilities of allocation to the healthy subgroup, there is substantial uncertainty around the boundary region in particular.  This uncertainty is taken into the account by the ASPR model but not by the other two-stage approaches.  Figure \ref{fig:post.pred} plots the raw data and also demonstrates the contours of posterior predictive density based on the MCMC samples of ASPR model. There seems no systematic discrepancy between the observations and the contours of posterior predictive density, suggesting the ASPR model fits the data well.  

\begin{table}
\caption{Posterior means and quantiles for component parameters in the ASPR model.\label{tbl:post.comp}}
\centering
{\footnotesize
\begin{tabular}{rrrrrr}
\toprule
& Mean & SD & 2.5\% & 50\% & 97.5\% \\ 
\hline
$\theta_{11}$& 237.52 & 4.54 & 228.06 & 237.67 & 245.93 \\  
$\theta_{12}$  & 2001.55 & 118.19 & 1751.22 & 2002.20 & 2219.74 \\ 
$\theta_{21}$  & 273.25 & 0.40 & 272.49 & 273.25 & 274.05 \\ 
$\theta_{12}$ & 3182.41 & 20.15 & 3141.49 & 3181.70 & 3223.60 \\ 
$\Sigma_{111}$& 829.19 & 134.00 & 590.10 & 822.11 & 1122.02 \\ 
$\Sigma_{112}$ & 19322.84 & 3255.32 & 13671.92 & 19100.50 & 26389.50 \\ 
$\Sigma_{122}$ & 508531.02 & 88244.72 & 367278.25 & 499645.00 & 695598.00 \\ 
$\Sigma_{211}$ & 96.78 & 6.52 & 85.01 & 96.32 & 110.15 \\ 
$\Sigma_{212}$ & 2174.75 & 239.50 & 1719.19 & 2168.40 & 2628.22 \\ 
$\Sigma_{222}$ & 235640.32 & 13202.36 & 211617.25 & 235185.00 & 262059.25 \\   
\bottomrule
\end{tabular}
}
\end{table}

\begin{figure}
\centering
\includegraphics[width=0.8\textwidth,angle=270]{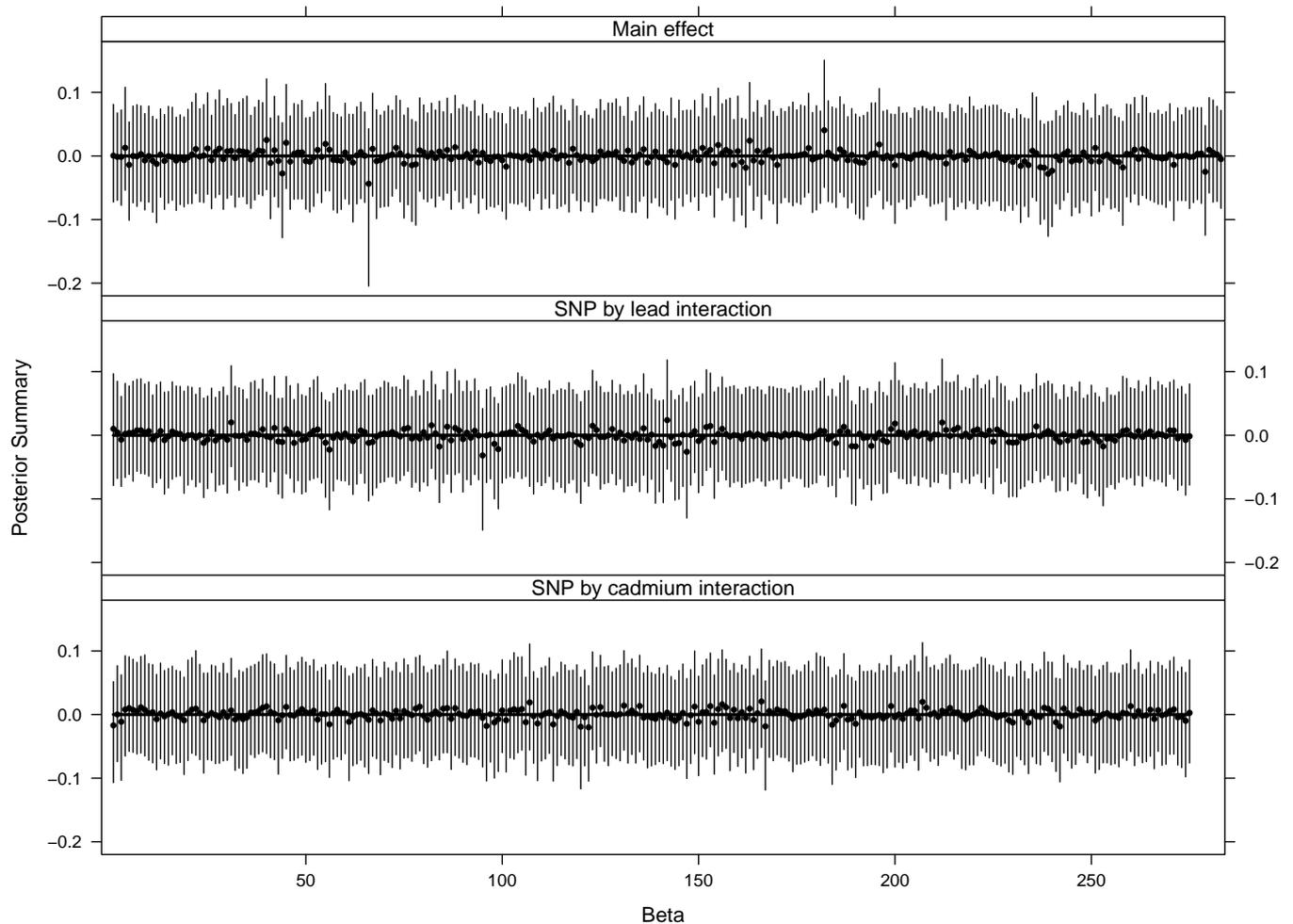}
  \caption{ Posterior mean and $90\%$ credible interval for coefficients $\bs{\beta}$ in ASPR model with MSP. The coefficients are illustrated for main effects and interactions respectively.}%; (c) posterior probability of minimum of PSA level at interval $[t_{j-1}, t_{j}]$ in SVM-OU}
   \label{fig:beta}
\end{figure}

Figure \ref{fig:beta} shows the posterior means and 90\% credible intervals for $\beta_j$, $j=1,\ldots,833$. Although all the credible intervals cover zero, the posterior mean of $\beta_{66}$ and $\beta_{182}$, given in Table \ref{tbl:snp_list}, stay clearly away from zero in the main effect (the top panel).  This result is also supported by Web Figure 1, which plots the histogram of $\tilde{\pi}_j=\sum_{g=1}^{G} I(|\beta^g_j|>\epsilon)/G$, the posterior probability that $j$th predictor have an effect based on G samples of $\beta_j$. When we take $\epsilon = 0.1$, $\tilde{\pi}_{66}$ stands out with the value $0.119$ and $\tilde{\pi}_{182}$ with the value $0.094$ while the majority of other posterior probabilities are around $0.05$. $\beta_{66}$ corresponds to SNP rs$3740564$ located in gene GRK5 and $\beta_{182}$ corresponds to SNP rs$10109780$ located in gene NAT1.  GRK5 is a member of the G protein-coupled receptor kinase family which is involved in regulating the activity levels of G protein-coupled receptors.  Polymorphisms in GRK5 have been previously linked to risk for heart failure in African Americans \cite{liggett2008grk5}.  The N-acetyltransferase genes (NAT1 and NAT2) are involved in the metabolism of xenobiotics.  NAT1 has been shown to be expressed in early placenta \cite{smelt2000expression}. We also analyzed the data using two-stage methods with the penalized logistic regressions and the indicators generated by maximum a posteriori method. The results are presented in Table \ref{tbl:snp_list}. Both penalized logistic regression methods identifies the predictor $X_{66}$ (SNP rs$3740564$ in GRK5) and  $X_{196}$ (SNP rs$7387461$ in NAT1).
For ASPR and penalized logistic regression with elastic net, additional $\beta$'s are interesting but not consistent across the three approaches. This may suggest that these are false positive results. 

\begin{table}
\caption{List of SNPs which have largest estimated SNP effects for different methods. \label{tbl:snp_list}}
\centering
{\footnotesize
\begin{tabular}{ccccccccccc}
\toprule
\multicolumn{3}{c}{ASPR-MSP}&&\multicolumn{3}{c}{Logit-Lasso}&&\multicolumn{3}{c}{Logit-ElasticNet}\\
\cline{1-3} \cline{5-7} \cline{9-11}
Coefficient& Estimate & Gene located && Coefficient& Estimate & Gene located &&Coefficient& Estimate & Gene located\\
\\
$\beta_{66}$&-0.044&GRK5 &&$\beta_{66}$&-0.100&GRK5 && $\beta_{66}$&-0.124&GRK5\\
$\beta_{182}$&0.040&NAT1 &&$\beta_{196}$&0.090&NAT1 && $\beta_{196}$&0.119&NAT1\\
$\beta_{378}$&-0.032&IGF1*Lead && && && $\beta_{339}$&-0.024& GRK5*Lead\\
$\beta_{239}$&-0.028&NOTCH1 && && &&  $\beta_{745}$&0.007&NAT1*Cadmium\\
$\beta_{44}$&-0.028&CYP2A6 && && && &&\\
\bottomrule
\end{tabular}
}
\end{table}

%\begin{figure}[h!]
%  \centering
%    \includegraphics[width=0.5\textwidth,angle=270]{figures/hist}
%  \caption{Posterior probability $\textsf{Pr}(\beta_j \neq 0)$ for genetic effects, environmental effects and their interactions. \label{fig:his}}
%\end{figure}

\section{Discussion}
\label{sec:Disc}
In this article, we propose an adverse subpopulation regression model for investigating the relationship among multiple quantitative outcomes and high-dimensional predictors. Unlike the traditional two-stage methods, the proposed method does not require dichotomizing the continuous outcomes into binary indicators and thus avoids information loss. { Two stage methods are outperformed } with smaller MSE and higher area under the ROC curve for variable selection as demonstrated by the simulation studies. The new model has been applied to examine the effect of gene and environment interaction on adverse pregnancy outcomes. The results suggest the gene GRK5 and NAT1 may influence the occurrence of low birth weight and preterm delivery.

{ Our focus is on defining a simple approach for assessing the impact of high-dimensional predictors on the risk of an adverse outcome when data consist of multiple quantitative traits.  By using a two component mixture model, we can use a binary response logistic regression model, a framework that is very familiar to epidemiologists, to characterize non-linear genetic and environment associations with potentially complex multivariate quantitative traits.  The proposed framework provides a parsimonious alternative to normal linear regression and logistic regressions based on preliminary categorization of quantitative traits, and should be able to detect associations that would not be detected with these methods.  The proposed ASPR framework has purposefully been chosen to be a simple and parsimonious model that is easy to interpret and is scalable to high-dimensional predictors.  For sake of simplicity and parsimony, we have avoided fully nonparametric Bayesian density regression models \cite{dunson2007bayesian} that allow unknown numbers of latent classes.  Although generalizations in such directions are conceptually straightforward, for each additional latent class, one introduces an additional $p$ regression coefficients and corresponding hyperparameters, and difficult issues in identifiability, label switching and computational complexity arise.  For our pregnancy outcome application, the ASPR model provide a good fit to the data, as illustrated by the posterior predictive density.}

\section{Supplementary Materials}
Web Appendix and Web Figure referenced in Sections \ref{sec:app} are available at the {\em Statistics in Medicine} website
{\tt \urlhttp onlinelibrary\urldot wiley\urldot com/journal/10.1002/(ISSN)1097-0258}.

\section*{Acknowledgments}
This work was supported by Award Number R01ES017436 from the National Institute of Environmental Health Sciences, and by funding from the National Institutes of Health (5P2O-RR020782-O3) and the U.S. Environmental Protection Agency (RD-83329301-0). The content is solely the responsibility of the authors and does not necessarily represent the official views of the National Institute of Environmental Health Sciences, the National Institutes of Health or the U.S. Environmental Protection Agency.

\bibliographystyle{wileyj} 
\bibliography{aspr}
                                                                             
\end{document}